\input harvmac
\input amssym

\def\Omega{\rho,\sigma,\nu  }

%% MACROS

\def\IL{\relax{\rm I\kern-.18em L}}
\def\IH{\relax{\rm I\kern-.18em H}}
\def\IR{\relax{\rm I\kern-.18em R}}
\def\IC{\relax\hbox{$\inbar\kern-.3em{\rm C}$}}
\def\IZ{\relax\ifmmode\mathchoice
{\hbox{\cmss Z\kern-.4em Z}}{\hbox{\cmss Z\kern-.4em Z}}
{\lower.9pt\hbox{\cmsss Z\kern-.4em Z}} {\lower1.2pt\hbox{\cmsss
Z\kern-.4em Z}}\else{\cmss Z\kern-.4em Z}\fi}
\def\ell_0{L_0}

\def\CN {{\cal N}}

%% MORE MACROS

\def\CN {{\cal N}}

\font\manual=manfnt \def\dbend{\lower3.5pt\hbox{\manual\char127}}

\def\IZ{\relax\ifmmode\mathchoice
{\hbox{\cmss Z\kern-.4em Z}}{\hbox{\cmss Z\kern-.4em Z}}
{\lower.9pt\hbox{\cmsss Z\kern-.4em Z}} {\lower1.2pt\hbox{\cmsss
Z\kern-.4em Z}}\else{\cmss Z\kern-.4em Z}\fi}
\def\half {{1\over 2}}

\def\rt2{\sqrt{2}}
\def\irt2{{1\over\sqrt{2}}}

%  \slashchar puts a slash through a character to represent contraction
%  with Dirac matrices. Use \not instead for negation of relations, and use
%  \hbar for hbar.
\def\slashchar#1{\setbox0=\hbox{$#1$}           % set a box for #1
   \dimen0=\wd0                                 % and get its size
   \setbox1=\hbox{/} \dimen1=\wd1               % get size of /
   \ifdim\dimen0>\dimen1                        % #1 is bigger
      \rlap{\hbox to \dimen0{\hfil/\hfil}}      % so center / in box
      #1                                        % and print #1
   \else                                        % / is bigger
      \rlap{\hbox to \dimen1{\hfil$#1$\hfil}}   % so center #1
      /                                         % and print /
   \fi}

%\GaiottoGF
\lref\GaiottoGF{
  D.~Gaiotto, A.~Strominger and X.~Yin,
  ``New connections between 4D and 5D black holes,''
  arXiv:hep-th/0503217.
  %%CITATION = HEP-TH 0503217;%%
}
%\GopakumarJQ
\lref\gv{
  R.~Gopakumar and C.~Vafa,
  ``M-theory and topological strings. II,''
  arXiv:hep-th/9812127.
  %%CITATION = HEP-TH 9812127;%%
}
%\DijkgraafXW
\lref\DijkgraafXW{
  R.~Dijkgraaf, G.~W.~Moore, E.~Verlinde and H.~Verlinde,
  ``Elliptic genera of symmetric products and second quantized strings,''
  Commun.\ Math.\ Phys.\  {\bf 185}, 197 (1997)
  [arXiv:hep-th/9608096].
  %%CITATION = HEP-TH 9608096;%%
} \lref\borch{R. E. Borcherds, "Automorphic forms on
$O_{s+2,2}(R)$ and infinite products" Invent. Math. {\bf 120}
(1995) 161.}

 \lref\KAWAI{
  T.~Kawai,
  ``$N=2$ heterotic string threshold correction, $K3$ surface and generalized
  Kac-Moody superalgebra,''
  Phys.\ Lett.\ B {\bf 372}, 59 (1996)
  [arXiv:hep-th/9512046].
  %%CITATION = HEP-TH 9512046;%%
}

\lref\DVV{
  R.~Dijkgraaf, E.~Verlinde and H.~Verlinde,
  ``Counting dyons in N = 4 string theory,''
  Nucl.\ Phys.\ B {\bf 484}, 543 (1997)
  [arXiv:hep-th/9607026].
  %%CITATION = HEP-TH 9607026;%%
}

%\AntoniadisZN
\lref\gava{
  I.~Antoniadis, E.~Gava, K.~S.~Narain and T.~R.~Taylor,
  %``N=2 type II heterotic duality and higher derivative F terms,''
  Nucl.\ Phys.\ B {\bf 455}, 109 (1995)
  [arXiv:hep-th/9507115].
  %%CITATION = HEP-TH 9507115;%%
}

\lref\BMPV{
  J.~C.~Breckenridge, R.~C.~Myers, A.~W.~Peet and C.~Vafa,
  ``D-branes and spinning black holes,''
  Phys.\ Lett.\ B {\bf 391}, 93 (1997)
  [arXiv:hep-th/9602065].
  %%CITATION = HEP-TH 9602065;%%
}

\lref\ascv{ A.~Strominger and C.~Vafa,
  ``Microscopic Origin of the Bekenstein-Hawking Entropy,''
  Phys.\ Lett.\ B {\bf 379}, 99 (1996)
  [arXiv:hep-th/9601029].}

%\LopesCardosoXF
\lref\LopesCardosoXF{
  G.~Lopes Cardoso, B.~de Wit, J.~Kappeli and T.~Mohaupt,
  ``Asymptotic degeneracy of dyonic N = 4 string states and black hole
  entropy,''
  JHEP {\bf 0412}, 075 (2004)
  [arXiv:hep-th/0412287].
  %%CITATION = HEP-TH 0412287;%%
}

%\draftmode

\newbox\tmpbox\setbox\tmpbox\hbox{\abstractfont }
\noblackbox
 \Title{\vbox{\baselineskip12pt\hbox to\wd\tmpbox{\hss
}}\hbox{hep-th/0505094 }} {\vbox{\centerline{Recounting Dyons in
$\CN=4$ String Theory}\medskip\centerline{} }}

\centerline{David Shih,\footnote{*}{Permanent address: Department of
Physics, Princeton University, Princeton, NJ 08544, USA.}~ Andrew
Strominger\footnote{**}{Permanent address: Jefferson Physical
Laboratory, Harvard University, Cambridge, MA 02138, USA.} and Xi
Yin** }
\smallskip\centerline{Center of Mathematical Sciences}
\centerline{ Zhejiang University, Hangzhou 310027 China}

\centerline{} \vskip.4in \centerline{\bf Abstract} {A recently
discovered relation between 4D and 5D black holes is used to
derive weighted BPS black hole degeneracies for  4D $\CN=4$ string
theory from the well-known 5D degeneracies. They are found to be
given by the Fourier coefficients of the unique weight 10
automorphic form of the modular group $Sp(2,\Bbb Z)$.   This
result agrees exactly with a conjecture made some years ago by
Dijkgraaf, Verlinde and Verlinde.

 } \vskip.2in

\vskip .2cm \Date{}

A general D0-D2-D4-D6 black hole in a 4D IIA string
compactification has an M-theory lift to a 5D black hole
configuration in a  multi-Taub-NUT geometry. This observation was
used in \GaiottoGF\ to derive a simple relation between 5D and 4D
BPS black hole degeneracies. For the case of $K3\times T^2$
compactification, corresponding to $\CN=4$ string theory, the
relevant 5D black holes were found in \refs{\ascv, \BMPV} and the
degeneracies are well known. In this paper we translate this into
an exact expression for the 4D degeneracies, which turn out to be
Fourier expansion coefficents of a well-studied weight 10
automorphic form $\Phi$ of the modular group of a genus 2 Riemman
surface \refs{\borch,\DVV}.

Almost a decade ago an inspired conjecture was made \DVV\ by
Dijkgraaf, Verlinde and Verlinde for the 4D degeneracies of
$\CN=4$ black holes, and this was shown to pass several
consistency checks. We will see that our analysis precisely
confirms their old conjecture.

$\CN=4$ string theory in four dimensions can be obtained from IIA
compactification on $K3\times T^2$. The duality group is
conjectured to be \eqn\dgr{SL(2;\Bbb Z)\times SO(6,22;\Bbb Z).}
The first factor may be described as an electromagnetic S-duality
which acts on electric charges $q_{e\Lambda}$ and magnetic charges
$q_m^\Lambda$, $\Lambda=0,..27$ transforming in the 28 of the
second factor. For the electric objects, we may take
\eqn\dsz{q_e=(q_0; q_A; q_{23}; q_j),} where $q_0$ is D0-charge,
$q_A, A=1,...22$ is $K3$-wrapped D2 charge, $q_{23}$ is
$K3$-wrapped D4 charge, and $q_i, i=24,...27$ are momentum and
winding modes of $K3\times S^1$-wrapped NS5 branes. The magnetic
objects are 24 types of D-branes which wrap $T^2\times ({\rm  K3
~cycle})$  and 4 types of F-string $T^2$ momentum/winding modes.

Now consider a black hole corresponding to a bound state of a
single D6 brane with  D0 charge $q_0$, $K3$-wrapped D2 charge
$q_A$, and  $T^2$-wrapped D2 charge $q^{23}$:\eqn\chargescorr{ q_m
= (1;q^A=0;q^{23};q^i=0),\quad q_e = (q_0; q_A;q_{23}=0;q_i=0) }
The duality invariant charge combinations are \eqn\chargeinv{
\half q_e^2 = \half C^{AB}q_A q_B,\quad \half q_m^2 = q^{23},\quad
q_e\cdot q_m = q_0 } where $C^{AB}$ is the intersection matrix on
$H^2(K3; \Bbb Z)$.

By lifting this to M-theory on Taub-NUT, it was argued in
\GaiottoGF\ that the BPS states of this system are the same as
those of a 5D black hole in a $K3\times T^2$ compactification,
with  $T^2$-wrapped M2 charge $\half q_m^2$, $K3$-wrapped M2
charge $q_A$ and angular momentum $J_L=q_0/2$. We now use one of
the compactification circles to interpret the configuration as IIA
on $K3\times S^1$ with $\half q_m^2$ F-strings winding $S^1$ and
$q_A$ D2-branes. T-dualizing the $S^1$ yields $q_A$ D3-branes
carrying momentum  $\half q_m^2$. This is then U-dual to a $Q_1$
D1 branes and $Q_5$ D5 branes on $K3\times S^1$ with
\eqn\QoneQfive{N \equiv Q_1Q_5 = {1\over2}q_e^2+1 } angular
momentum\foot{One should keep in mind that $J_L$ is half the
R-charge $F_L$ \BMPV, and is hence takes values in ${1\over2}\Bbb
Z$.} \eqn\angular{ J_L = {1\over2}q_e\cdot q_m } and left-moving
momentum along the $S^1$: \eqn\momentum{ \ell_0 = {1\over2}q_m^2
.} Hence, with the above relations between parameters,  according
to \GaiottoGF\ the 4D degeneracy of states with charges
\chargescorr\ and 5D degeneracies are related by
\eqn\crfd{d_4(1;0;q^{23};0|q_0;q_A;0;0)=(-1)^{q_0}d_5\left(q^{23},q_A;{q_0
\over 2}\right).} The extra factor of $(-1)^{q_0}$ comes from the
extra insertion of $(-1)^{2J_L}$ in the definition of the 5D
index. Since the degeneracies are U-dual we may also
write\foot{Note that $d_n$ denotes fixed-charge degeneracies and
does not involve a sum over U-duality orbits. }
\eqn\donedfivemicro{d_4( q_m^2, q_e^2,  q_e\cdot q_m )=
(-1)^{2J_L}d_5(\ell_0,N,J_L)=(-1)^{q_e\cdot q_m}d_5\left(\half
q_m^2,\half q_e^2+1, \half q_e\cdot q_m\right).} Here and
elsewhere in this paper by ``degeneracies," in a slight abuse of
language, we mean the number of  bosons minus the number of
fermions of a given charge, and the center-of-mass multiplet is
factored out.

Of course these microscopic BPS degeneracies $d_5$  of the D1-D5
system are well known \refs{\ascv,\BMPV}. Their main contribution
comes from the coefficients in the Fourier expansion of the
elliptic genus of ${\rm Hilb}^N(K3)$: \eqn\entropyhilbert{
\chi_N(\rho,\nu) = \sum_{\ell_0,J_L} d'_5(\ell_0, N,J_L)e^{2\pi
i(\ell_0\rho +2J_L\nu )} } It is shown in \DijkgraafXW\ that the
weighted sum of the elliptic genera has a product representation:
\eqn\generaprod{ \sum_{N\ge 0} \chi_N(\rho,\nu) e^{2\pi iN\sigma}
= {1\over \Phi'(\rho,\sigma,\nu)} } where $\Phi'$ is given by
\eqn\phiprime{\Phi'(\rho,\sigma,\nu) = \prod_{k\geq 0, l>0, m\in
{\Bbb Z}} (1-e^{2\pi i (k\rho+l\sigma+m\nu)})^{c(4kl-m^2)}, } with
$c(4k-m^2)=d'_5(k,1,m)$ the elliptic genus coefficients for a
single $K3$ as given in \KAWAI.\foot{Note $c(-1)=2$, $c(0)=20$,
and $c(n)=0$ for $n\leq -2$.}

Equation \generaprod\ is the generating function for BPS states of
CFTs on ${\rm Hilb}^N(K3)$ in the D5 worldvolume. However it does
not quite give the degeneracies needed in \donedfivemicro\ because
it leaves out the decoupled contribution from the elliptic genus
of a single fivebrane.  This remains even when $N=0$ and there are
no D1 branes at all. (By U-duality, we are free to view the system
as a single fivebrane and $N$ D1 branes.) Using the U-dual
relation of a $K3$ -wrapped D5 brane to a fundamental heterotic
string, the elliptic genus, not including the center of mass
contribution, is \refs{\gava,\gv} \eqn\mismmz{
Z_0(\nu,\rho)=(e^{\pi i\nu}-e^{-\pi i\nu})^{-2} e^{-2\pi i
\rho}\prod_{n\geq 1}(1-e^{2\pi i(n\rho+\nu)} )^{-2} (1-e^{2\pi
i(n\rho-\nu)} )^{-2} (1-e^{2\pi i n \rho})^{-20}. } This shifts
$\Phi'$ to \eqn\phidiffprime{{1 \over\Phi'(\Omega)} \to
{Z_0(\nu,\rho) \over\Phi'(\Omega)} = {e^{2\pi i \sigma}\over
\Phi(\Omega)} %=
%\sum_{L_0,N,J_L}d_5(L_0,N,J_L)e^{2\pi i( \rho L_0+\sigma
%(N-1)+2\nu J_L)}
} where $\Phi(\Omega)$ has a product representation \eqn\phiprod{
\Phi(\Omega) = e^{2\pi i (\rho+\sigma+\nu)} \prod_{(k,l,m)>0}
\left( 1 - e^{2\pi i (k\rho+l\sigma+m\nu)} \right)^{c(4kl-m^2)} }
where $(k,l,m)>0$ means that $k,l\geq 0$, $m\in {\Bbb Z}$ and in
the case $k=l=0$, the product is only over $m<0$. $\Phi(\Omega)$
is the unique automorphic form of weight 10 of the modular group
$Sp(2,\Bbb Z)$ and was studied in \borch. The 5D BPS degeneracies
are then the Fourier coefficients in
 \eqn\identitycheck{
\sum_{\ell_0,N,J_L}d_5(\ell_0,N,J_L)e^{2\pi
i(\ell_0\rho+(N-1)\sigma+2J_L\nu)} = {1\over\Phi(\rho,\sigma,\nu)}
.} Inserting the 4D-5D relation \donedfivemicro, \identitycheck\
agrees with the formula proposed in \DVV\ for the microscopic
degeneracy of BPS black holes of $\CN=4$ string theory -- up to an
overall factor of $(-1)^{q_e\cdot q_m}$.\foot{Note that the
formula of \DVV\ was manifestly invariant under the duality group
\dgr, and the extra factor of $(-1)^{q_e\cdot q_m}$ does not spoil
this. Invariance of $(-1)^{q_e\cdot q_m}$ under $SO(6,22;{\Bbb
Z})$ follows by construction, and invariance under $SL(2;{\Bbb
Z})$ follows from the S-duality transformations of the charges.}

\bigskip\bigskip\bigskip

\centerline{\bf Acknowledgments} This work was supported in part
by DOE grant DEFG02-91ER-40654.  We are grateful to Allan Adams
and Greg Moore for useful conversations and correspondence.

 \listrefs
\end